\documentclass[runningheads]{cl2emult}

\usepackage{makeidx}  
\usepackage{graphicx} 
\usepackage{subeqnar} 
\usepackage{multicol} 
\usepackage{cropmark} 
\usepackage{eso}      
\makeindex            

\begin{document}
\title*{ Tessellation Reconstruction Techniques}
%
\titlerunning{Tessellation Reconstruction Techniques}
%
\author{Rien van de Weygaert
\and Willem Schaap}
\authorrunning{Rien van de Weygaert \& Willem Schaap}
%
%
\institute{Kapteyn Astronomical Institute, University of Groningen, the Netherlands}
\maketitle              

\begin{abstract}
The application of Voronoi and Delaunay tessellation based methods for 
reconstructing continuous fields from discretely sampled data sets is 
discussed. The succesfull operation as ``multidimensional interpolation'' 
method is corroborated through their ability to reproduce even intricate 
statistical aspects of the analytical predictions of perturbation theory 
of cosmic velocity field evolution. The newly developed and fully 
self-adaptive technique for density field estimation by means of 
Delaunay tessellations, basically exploiting their ``minimum triangulation'' 
quality, is shown to succesfully reproduce the morphology of the foamlike 
structure in N-body simulations of cosmic structure formation. The 
full hierarchy of structure is implicitly and directly reproduced at every 
spatial resolution scale present in the particle distribution, at the 
same time automatically and sharply rendering its characteristic 
anisotropic filamentary and wall-like features.  
\end{abstract}
\section{Continuous Fields sampled by Discrete Data Sets}
Astronomical observations, physical experiments as well as computer
simulations often involve discrete data sets supposed to represent a
fair sample of an underlying smooth and continuous field. 
Reconstructing the underlying fields from a set of irregularly 
sampled data is therefore a recurring key issue in operations on astronomical 
data sets. 

Within the context of the reconstruction issue, we may distinguish two 
basically distinct situations. One is that of a specific 
continuous field whose values have been measured at a set of discrete 
locations. A typical example, of a cosmological nature, concerns the 
sampling of the global cosmic matter flow involved in the build up of 
structure in the Universe. The measured peculiar 
velocities of galaxies are supposed to be a fair reflection of 
the underlying cosmic flow. The reconstruction problem may then 
be described as  an issue of ``Multidimensional Interpolation''. Evidently, 
a myriad of astronomical studies involve related  such issues. A second 
class involves the issue of estimating the underlying continuous intensity 
or density field from a point process supposedly representing a fair 
sampling of this field. The fact that the sampling point process itself is a 
reflection of the underlying field forms an extra complication in the case 
of these ``Optimal Density Estimation'' problems.  
\begin{figure*}
  \includegraphics*[width=\hsize,keepaspectratio]{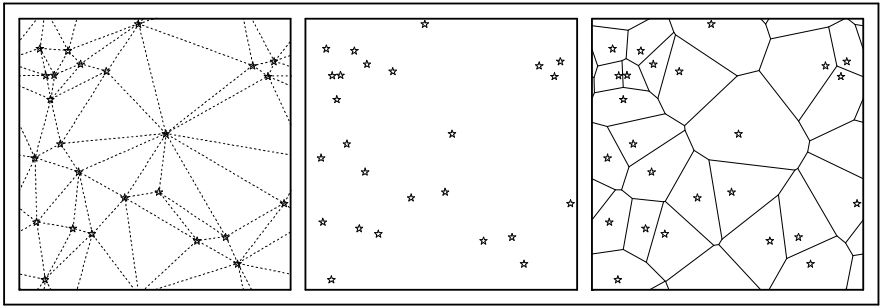}
  \caption{\small The Delaunay triangulation (left frame) and Voronoi 
tessellation (right frame) of a distribution of 25 nuclei (stars) in a 
square (central panel). Periodic boundary conditions are assumed.}
\end{figure*}
Conventional reconstruction methods are usually plagued by one or more 
artefacts. Firstly, they often involve estimates at a finite number 
of locations, usually confined to a grid. Optimal field 
estimators should provide a prescription for the value of that field throughout 
the whole sampling volume. Conventional schemes usually restrict 
themselves to estimates at a finite number of positions. 
For most practical purposes, a further disadvantage of almost all 
conventional methods is their insensitivity and inflexibility to the 
sampling point process. This leads to a far from optimal performance in both 
high density and low density regions, which often is dealt with by rather 
artificial and ad hoc means. 

In particular in situations of highly non-uniform distributions 
conventional methods tend to obscure various interesting and 
relevant aspects present in the data. An illuminating example is the case 
of the galaxy distribution and the cosmic matter density field. 
Redshift surveys as well as computer simulations of structure 
formation reveal salient anisotropic features like filaments  
and walls, extended along one or two directions while 
compact in the other(s). In addition, the density fields 
display hierarchical structure of varying contrasts over a 
large range of scales. Ideally sampled by the data points, 
appropriate field reconstructions should be set solely and automatically 
by the sampling point distribution itself. 
The commonly used methods, involving artificial filtering through grid size 
or other smoothing kernels (e.g. Gaussian filter) 
mostly fail to achieve an optimal result and are neither able to 
reproduce the tenuous anisotropic structures seen in the point process, 
nor a faithful rendering of the full structural hierarchy. 
A final, usually concealed, yet fundamental aspect of field reconstruction 
turns out to be of key significance for the definition of the presented tessellation 
methods. Most well-known methods implicitly yield mass-weighted averages. This 
renders the comparison with volume-weighted analytical quantities 
far from trivial. Tessellation methods represent an elegant solution to 
this point.
\section{Volume-averaged quantities}
Given a discrete set of field values $f({\bf x})$ measured at 
${\rm N}$ locations, the usual 
procedure of field reconstruction consists of smoothing the 
measured discrete field values by some filter function. A {\it mass-weighted} 
field $f_{mass}({\bf x})$, \begin{equation} f_{mass}({\bf x}_0) \equiv {\displaystyle \int 
d{\bf x}\,f({\bf x}) \,\rho({\bf x}) W_M({\bf x},{\bf x}_0) \over 
\displaystyle \int d{\bf x}\,\rho({\bf x}) W_M({\bf x},{\bf x}_0)}\,
\end{equation}
with $W_M({\bf x},{\bf x}_0)$ a filter function, is almost 
without exception the conventionally employed filtering scheme.
An example of this is probably one of the most frequently 
applied class of filtering schemes, involving the interpolation of 
field values at random sampling (galaxy) locations to those 
at regular grid locations, weighing the contribution by each sampling 
point by the filter function value. However, the presence of the 
the extra mass-weighting density field factor $\rho$ introduces 
considerable technical repercussions. Analytical treatments of the 
corresponding physics -- in particular in the case of a perturbation 
analysis -- is often almost exclusively limited to 
{\it volume-weighted} filtered fields $f_{vol}({\bf x})$,
\begin{equation} f_{vol}({\bf x}_0) \equiv {\displaystyle
\,\int d{\bf x}\,f({\bf x}) W_V({\bf x},{\bf x}_0) \over 
\displaystyle \int d{\bf x}\,W_V({\bf x},{\bf x}_0)}\,, 
\end{equation} with $W_V({\bf x},{\bf x}_0)$ the applied weight functionx. In general 
the properties and behaviour of volume-weighted and mass-weighted quantities 
will be fundamentally different. A succesfull comparison and assessment of 
observational or numerical results with relevant physical theory will 
therefore essentially only be sensible if we have a reliable 
numerical estimators of volume-averaged quantities. 

In the ideal but unrealistic situation the underlying continuous field 
would be reproduced from a set of field values sampled at an infinite 
number points and subsequently volume filtered with a filter function $W_V$ whose 
filter radius is infinitely small. Extrapolating from this premise, a good 
approximation of volume average quantities is obtained by volume averaging over 
quantities that were mass filtered onto a extremely fine-mazed grid with, 
in comparison, a very small scale for the mass weighting filter function.
In most practical circumstances, however, the above prescription resorts to 
coarse grids, leading to field estimates whose quality is not readily 
appreciated. Alternatively, we may specifically pursue the fact that the 
corresponding sampling density field can be described as the sum of delta functions 
at each sample location. Combining this with the asymptotic limit of 
applying a volume weighted filter with an infinitely small filter scale, we find 
the first-step field estimate $f_1({\bf x})$ (Eq.~(3)) through ordering  
the locations $i$ by increasing distance to 
${\bf x}_0$ and thus by decreasing value of $w_i$. It is easy to see (Bernardeau \& van 
de Weygaert 1996) that 
this defines a field $f({\bf x})$ in which the field value at every location in space 
acquires the value of the field at the closest point of the discrete field sample,  
\begin{equation}f_{1}({\bf x}_0)\,=\,{\displaystyle \sum_i w_i f({\bf x}_i) 
\over \displaystyle \sum_i w_i}
\,=\, {{\displaystyle f({\bf x_1})+ \sum_{i=2}^N {w_i \over w_1} f({\bf x}_i}) 
\over \displaystyle 1 + \sum_{i=2}^N {w_i \over w_1}}\quad \longrightarrow 
\quad f({\bf x}_1)\end{equation}
\section{Voronoi and Delaunay Tessellation Interpolation}
The procedure described above implies nothing else than the concept of
the {\it Voronoi tessellation} (Fig.~1). Such a tessellation consists of 
a space-filling network of mutually disjunt convex polyhedral cells, 
the {\it Voronoi polyhedra}, each of which delimits the part of space 
that is closer to the defining point in the discrete point sample 
set than to any of the other sample points (see Icke \& Van de Weygaert 1987, 
and Van de Weygaert 1991, 1994, for extensive descriptions and references). 
\begin{figure*}
\sidecaption\includegraphics*[height=6.0cm,keepaspectratio]{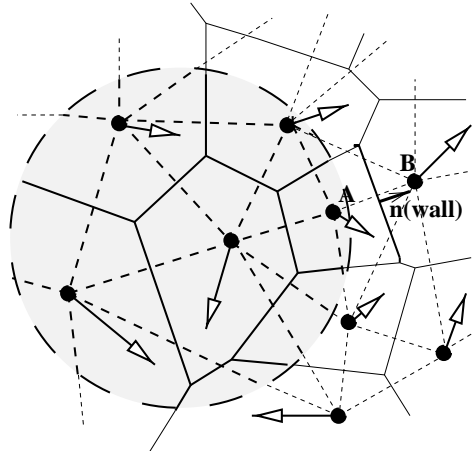} 
\caption{Voronoi and Delaunay tessellations of a 2D set of particles 
(filled circles). The solid lines form the Voronoi tessellation, the 
dashed lines the Delaunay tessellation. We also indicated a normal
vector {\bf n} of the wall separating the points ${\bf A}$ and ${\bf
B}$. The grey circle represents the area in which one determines 
the top-hat filtered volume average of the velocity gradients. Courtesy: 
F. Bernardeau (see Bernardeau et al. 1997)}
\end{figure*}
The {\it Voronoi method} defined in this way is based on the assumption 
that the field is uniform within each Voronoi cell of the tessellations, such that 
the field value throughout each of the Voronoi polyhedra is equal to that of the 
sample point of the tessellation (see Fig. 2 for a summarizing illustration of 
the method). Basically, the Voronoi cells are the 
multidimensional generalization of the bins in a 1-D zeroth-order 
interpolation scheme in which the function value is supposed to be constant 
within the interval centered on the sampling points, their value equal to that of 
the respective sample point values. Bernardeau \& Van de Weygaert (1996) showed 
its great performance in the case of evaluating cosmic velocity fields. 
They showed the method to be of particular benefit when considering 
the gradient fields of discretely sampled fields rendered by the Voronoi method. 
In such situations, the Voronoi method yields a spatial geometry in which non-zero 
values of the field gradients are solely localized to the (polygonal) Voronoi walls. 
For the specific case of the window function for the volume filtering being 
a top-hat filter, the subsequent computation of the volume averages of the 
field gradients consist of a relatively simple sum of the values of those 
field gradients in each of the tessellation walls $k$ intersected or inside the 
filter sphere weighted by the surface area $A_k$ of the part of the wall 
located within the sphere. 

The specific application to the statistics of the velocity divergence field 
$\nabla \cdot {\bf v}$ was shown to be very succesfull, corroborating the 
prediction of analytical perturbation theory considerations from a set of 
N-body simulations of structure formation (Bernardeau \& Van de Weygaert 1996).  
Notwithstanding its virtues, the Voronoi method evidently represents an 
artificial situation in which the reconstructed fields are emphatically 
discontinous. Moreover, it cannot be applied to filter radii 
that are smaller than the average Voronoi wall distance. Below those
scales the probability that a randomly placed filter sphere does not 
contain or intersect any Voronoi wall gets prohibitively high, and 
yielding unrealistic zero values for the field gradients. 

In the one-dimensional situation a first-order improvement concerns the 
linear interpolation between the sampling points, leading to a fully 
continuous field. The natural extension to a multidimensional linear
interpolation interval then immediately implies the corresponding
Delaunay tessellation (Delone 1934). This tessellation (Fig.~1)
consists of a volume-covering tiling of space into tetrahedra (in 3-D,
triangles in 2-D, etc.) whose vertices are formed by four specific
points in the dataset. The four points are uniquely selected such that
their circumscribing sphere does not contain any of the other
datapoints. The Voronoi and Delaunay tessellation are intimately
related, being each others dual in that the centre of each Delaunay
tetrahedron's circumsphere is a vertex of the Voronoi cells of each of
the four defining points, and conversely each Voronoi cell nucleus a
Delaunay vertex (see Fig.~1). The ``minimum triangulation'' property
of the Delaunay tessellation has in fact been well-known and
abundantly applied in, amongst others, surface rendering applications
such as geographical mapping and various computer imaging algorithms.
Consider a set of ${\rm N}$ discrete datapoints in a finite region of 
M-dimensional space. Having at one's disposal the field values at each of 
the $(1\!+\!{\rm M})$ Delaunay vertices ${\bf x}_0, \ldots, 
{\bf x}_{\rm M}$, at each location ${\bf x}$ in the interior of a Delaunay M-dimensional 
tetrahedron the linear interpolation field value is determined by 
the estimated constant field gradient within the 
tetrahedron, $\,\,f({\bf x})~=~f({\bf x}_0)~+~(\nabla f)|_{\rm Del} \!\!\!\cdot 
({\bf x}- {\bf x}_0)\,\,\,$. Given the $(1\!+\!{\rm M})$ field values  $f({\bf x}_i)$, 
the value of the ${\rm M}$ components of the 
Delaunay field gradient $~(\nabla f)|_{\rm Del}$ can be computed 
straightforwardly by solving this linear relation for each of the ${\rm M}$ points 
${\bf x}_1, \ldots, {\bf x}_{\rm M}$. 

This multidimensional {\it Delaunay procedure} of 
linear interpolation was introduced and described by Bernardeau \& Van de 
Weygaert (1996) in the context of defining procedures 
for volume-weighted estimates of cosmic velocity fields. As they specifically 
focussed on the statistics of the velocity field gradients, in essence they 
defined a velocity gradient field of constant gradient values within each 
individual Delaunay tetrahedron, its value set by the measured velocity 
values at each of the 4 Delaunay vertices. They showed the superior 
performance of the first-order Delaunay estimator in reproducing 
analytical predictions of gravitational instability perturbation theory. 
The virtues and promise of the {\it Delaunay method} is in particularly 
underlined by its ability to unequivocally estimate the 
cosmic density parameter $\Omega$ on the basis of the particle velocities 
in the various N-body simulations of cosmic structure formation (Bernardeau 
etal. 1997). 
\begin{figure*}
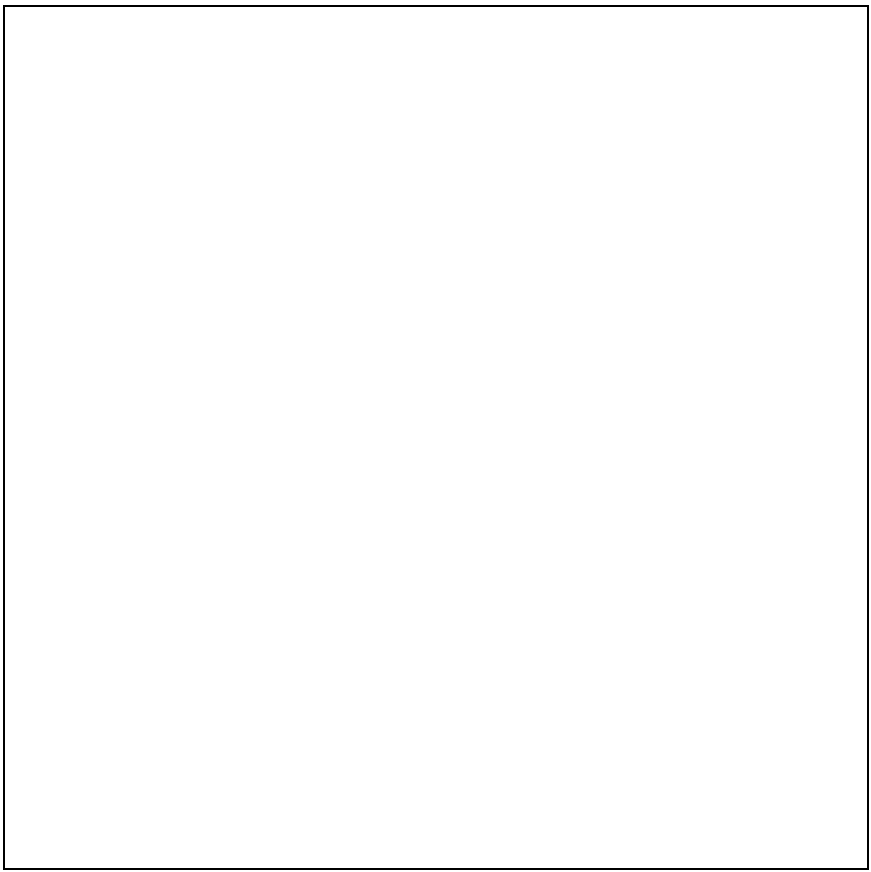

  \mpicplace{11.5 cm}{11.5 cm}
  \caption{\small A 9-frame mosaic comparing
    the performance of the Delaunay density estimating technique with
    a conventional grid-based TSC method in analyzing a cosmological
    N-body simulation. Left column: the particle distribution in a
    $10h^{-1}\hbox{Mpc}$ wide central slice through the simulation
    box. Central column: the corresponding Delaunay density field
    reconstruction. Right column: the TSC rendered density field
    reconstruction. The density grey scale is logarithmic, $\delta\rho/\rho=0-2400$.}
\end{figure*}
\section{Delaunay Density Field Estimation}
The one factor complicating a trivial and direct implementation of
above procedure in the case of density (intensity) field estimates is
the fact that the number density of data points itself is the measure
of the underlying density field value. We therefore cannot start with 
directly available field estimates at each datapoint. Instead, we need to define 
appropriate estimates from the point set itself. Most suggestive would be to base
the estimate of the density field at the location ${\bf x}_i$ of each
point on the inverse of the volume ${\rm V_{Vor,i}}$ of its Voronoi
cell, $\rho({\bf x}_i)\!=\!{\rm m}/{\rm V_{Vor,i}}$. Note that in this
we take every datapoint to represent an equal amount of mass ${\rm m}$.
The resulting field estimates are then intended as input for the above
Delaunay interpolation procedure. However, one can demonstrate that
integration over the resulting density field would yield a different
mass than the one represented by the set of sample points (see Schaap
\& Van de Weygaert 2000a,b for a more specific and detailed
discussion).  Instead, mass conservation is naturally guaranteed when
the density estimate is based on the inverse of the volume ${\rm
  W_{Vor,i}}$ of the ``contiguous'' Voronoi cell of each datapoint,
$\rho({\bf x}_i)\!\propto\!1/{\rm W_{Vor,i}}$. The ``contiguous''
Voronoi cell of a point is the cell consisting of the agglomerate of
all $K$ Delaunay tetrahedra containing point $i$ as one of its
vertices, whose volume ${\rm W_{Vor,i}}=\sum_{j=1}^K {\rm V_{Del,j}}$
is the sum of the volumes ${\rm V_{Del,j}}$ of each of the $K$
Delaunay tetrahedra. Properly normalizing the mass contained
in the reconstructed density field, taking into account the fact that
each Delaunay tetrahedron is invoked in the density estimate at
$1\!+\!{\rm M}$ locations, we find at each datapoint the following
density estimate, $\rho({\bf x}_i)={\rm{m\,(1\!+\!M)}}/{\rm W_{Vor,i}}$. 
Having computed these density estimates, we subsequently proceed to
determine the complete volume-covering density field reconstruction
through the linear interpolation procedure outlined above.

The outstanding performance of our {Delaunay Density Estimtor} is 
illustrated by Figure~3 (see Schaap \& Van de Weygaert 2000a), in which 
its performance on an N-body simulation of cosmic structure formation 
is compared with that of a conventional grid-based TSC technique.
Cosmological N-body simulations provide an ideal template for
illustrating the virtues of our method. They tend to contain a large
variety of objects, with diverse morphologies, a large reach of
densities, spanning over a vast range of scales. They display low
density regions, sparsely filled with particles, as well as highly
dense and compact clumps, represented by a large number of particles.
Moderate density regions typically include strongly anisotropic
structures such as filaments and walls. 
A comparison of the lefthand and righthand columns with the central
column, i.e.~the Delaunay estimated density fields, reveals the
striking improvement rendered by our new procedure. Going down from
the top to the bottom in the central column, we observe seemingly
comparable levels of resolved detail. The self-adaptive skills of the
Delaunay reconstruction evidently succeed in outlining the full
hierarchy of structure present in the particle distribution, at every
spatial scale represented in the simulation. The contrast with the
achievements of the fixed grid TSC method in the righthand column is
striking, in particular when focus tunes in on the finer structures.
The central cluster appears to be a mere featureless blob! In
addition, low density regions are rendered as slowly varying regions
at moderately low values.  This realistic conduct should be set off
against the erratic behaviour of the TSC reconstructions, plagued by
annoying shot-noise effects.
Figure~3 also bears witness to the additional success of the Delaunay 
Estimator in reproducing sharp, edgy and clumpy
filamentary and wall-like features. Automatically it resolves the fine
details of their anisotropic geometry, seemlessly coupling sharp
contrasts along one or two compact directions with the mildy varying
density values along the extended direction(s). Moreover, it also
manages to deal succesfully with the substructures residing within
these structures. The well-known poor operation of e.g.~the TSC method
is clearly borne out by the central righthand frame. Its fixed and
inflexible ``filtering'' characteristics tend to blur the finer
aspects of such anisotropic structures. Such methods are therefore
unsuited for an objective and unbiased scrutiny of the foamlike
geometry which so pre-eminently figures in both the observed galaxy
distribution as well as in the matter distribution in most viable
models of structure formation.

Evidently, unlike artificial tailor-made methods, the {\it Delaunay 
Density Estimator} is sensitive in a fully self-consistent and 
self-adaptive fashion to intrinsically important structural elements. This 
provides ample arguments for its promise in great many astrophysical 
environments. 
\section*{Acknowledgments}
We thank F.~Bernardeau for substantial contributions towards 
instigating this line of research.

\clearpage
\addcontentsline{toc}{section}{Index}
\flushbottom
\printindex


\begin{thebibliography}{7}
\addcontentsline{toc}{section}{References}
\bibitem{berwey96} Bernardeau, F., van de Weygaert, R. (1996). MNRAS {\bf 279}, 693--711
\bibitem{berel97} Bernardeau, F., van de Weygaert, R., Hivon, E., Bouchet, F. 
(1997). MNRAS {\bf 290}, 566-576
\bibitem{del34} Delone, B.N. (1934). Bull.~Acad.~Sci.~USSR: Classe Sci.~Mat. 
{\bf 7}, 793
\bibitem{ickwey87} Icke, V., Van de Weygaert, R. (1987). 
Astron.~Astrophys. {\bf 184}, 1--9
\bibitem{schwey2000a} Schaap, W., Van de Weygaert, R. (2000a). Astron.~Astrophys. 
(submitted).
\bibitem{schwey2000b} Schaap, W., Van de Weygaert, R. (2000b). Astron.~Astrophys. 
(in prep). 
\bibitem{wey91} Van de Weygaert, R. (1991) Ph.D. thesis, Leiden University, 1991
\bibitem{wey94} Van de Weygaert, R. (1994) Astron.~Astrophys. {\bf 283}, 361--406
\end{thebibliography}
\end{document}